\documentclass[prd,
superscriptaddress,
tightenlines,nofootinbib,
  eqsecnum
  ]{revtex4}

\usepackage{amsmath}
\usepackage{amsfonts}
\usepackage{amssymb}
\usepackage{bm}
\usepackage{comment}
\usepackage{hyperref}
\usepackage{mathrsfs}
\usepackage{graphicx}

\usepackage{ulem}
\usepackage[usenames]{color}

\newcommand{\be}{\begin{eqnarray}}
\newcommand{\ee}{\end{eqnarray}}
\newcommand{\bdm}{\begin{displaymath}}
\newcommand{\edm}{\end{displaymath}}
\newcommand{\ds}{\displaystyle}
\newcommand{\ba}{\begin{array}}
\newcommand{\ea}{\end{array}}

\newcommand{\K}{{\bf k}}
\newcommand{\Q}{{\bf q}}

\definecolor{darkgreen}{rgb}{0,0.5,0}

\hypersetup{
    bookmarks=true,         
    unicode=false,          
    pdftoolbar=true,        
    pdfmenubar=true,        
    pdffitwindow=false,     
    pdfstartview={FitH},    
    pdftitle={Energy logs}, 
    pdfauthor={Author},     
    pdfsubject={Subject},   
    pdfcreator={Creator},   
    pdfproducer={Producer}, 
    pdfkeywords={keyword1} {key2} {key3}, 
    pdfnewwindow=true,      
    colorlinks=true,       
    linkcolor=red,          
    citecolor=cyan,        
    filecolor=magenta,      
    urlcolor=darkgreen,           
    linktocpage=true
}

\allowdisplaybreaks

\DeclareSymbolFontAlphabet{\mathrsfs}{rsfs}
\DeclareMathAlphabet{\mathcal}{OMS}{cmsy}{m}{n}

\newcommand{\dd}{\mathrm{d}}
\newcommand{\di}{\mathrm{i}} 
\newcommand{\de}{\mathrm{e}} 
\newcommand{\beq}{\begin{equation}}
\newcommand{\eeq}{\end{equation}}

\newcounter{theorem} 

\begin{document}

\title{Logarithmic tail contributions to the energy function of circular compact binaries}

\author{Luc Blanchet}\email{luc.blanchet@iap.fr}
\affiliation{$\mathcal{G}\mathbb{R}\varepsilon{\mathbb{C}}\mathcal{O}$, Institut d'Astrophysique de Paris,\\ UMR 7095, CNRS, Sorbonne Universit{\'e},\\ 98\textsuperscript{bis} boulevard Arago, 75014 Paris, France}

\author{Stefano Foffa}\email{stefano.foffa@unige.ch}
\affiliation{D\'epartement de Physique Th\'eorique and Center for Astroparticle Physics, Universit\'e de 
             Gen\`eve, CH-1211 Geneva, Switzerland}

\author{Fran\c{c}ois Larrouturou}\email{francois.larrouturou@iap.fr}
\affiliation{$\mathcal{G}\mathbb{R}\varepsilon{\mathbb{C}}\mathcal{O}$, Institut d'Astrophysique de Paris,\\ UMR 7095, CNRS, Sorbonne Universit{\'e},\\ 98\textsuperscript{bis} boulevard Arago, 75014 Paris, France}

\author{Riccardo Sturani}\email{riccardo@iip.ufrn.br}
\affiliation{International Institute of Physics, Universidade Federal do Rio Grande do Norte, Campus Universit\'ario, Lagoa Nova, Natal-RN 59078-970, Brazil}

\date{\today}

\begin{abstract}
We combine different techniques to extract information about the logarithmic contributions to the two-body conservative dynamics within the post-Newtonian (PN) approximation of General Relativity. The logarithms come from the conservative part of non linear gravitational-wave tails and their iterations. Explicit, original expressions are found for conservative dynamics logarithmic tail terms up to 6PN order by adopting both traditional PN calculations and effective field theory (EFT) methods. We also determine all logarithmic terms at 7PN order, fixing a sub-leading logarithm from a tail-of-tail-of-tail process by comparison with self-force (SF) results. Moreover, we use renormalization group techniques to obtain the leading logarithmic terms to generic power $n$, appearing at $(3n+1)$PN order, and we resum the infinite series in a closed form. Half-integer PN orders enter the conservative dynamics starting at 5.5PN, but they do not generate logarithmic contributions up to next-to-next-to-leading order included. We nevertheless present their contribution at leading order in the small mass ratio limit.
\end{abstract}

\keywords{classical general relativity, coalescing binaries, post-Newtonian expansion, radiation reaction}

\pacs{04.20.-q,04.25.Nx,04.30.Db}


\maketitle

\section{Motivations and overview} 
\label{sec:intro}
The post-Newtonian (PN) approximation to General Relativity (GR) has been a largely successful framework to perturbatively solve Einstein's equations, widely adopted and approached with a variety of
methods, see Refs.~\cite{BuonSathya15, BlanchetLR, Goldberger:2007hy, FSreview, Porto:2016pyg} for recent reviews. Among the most important methods we mention the ADM Hamiltonian approach~\cite{S85, DS85, St11rev}, the Multipolar-post-Minkowskian framework with PN matching (MPM-PN)~\cite{BD86, B98mult}, the direct integration of the relaxed field equations (DIRE)~\cite{WW96}, the surface-integral approach~\cite{IFA00} and the Effective Field Theory (EFT) approach pioneered by Ref.~\cite{GR06}. In particular we want to highlight here the great synergies existing today between the EFT approach and more traditional PN methods.

Within the two-body dynamics, we focus in the present work on \emph{tail} processes, which arise from the back-scattering of gravitational waves (GW) off the quasi-static curvature sourced by the total mass of the binary system. Tail effects are known from a long time (see \emph{e.g.}~\cite{BoR66, Th80}) but were first identified and investigated in the present context in~\cite{BD88, BD92, B93}. They are present in both the conservative and dissipative sectors of the theory. The conservative tail effect at 4PN order~\cite{FStail, GLPR16} has been recently fully incorporated into the 4PN equations of motion using the ADM Hamiltonian method~\cite{JaraS13, DJS14, JaraS15, DJS16}, the Fokker Lagrangian in harmonic coordinates~\cite{BBBFMa, BBBFMb, BBBFMc, MBBF17} and the EFT approach~\cite{FS4PN, FMSS16, FS19, FPRS19}. Moreover the leading and next-to-leading logarithmic tail terms in the energy function of compact binaries on circular orbits have been derived~\cite{BDLW10b, D10sf, LBW12, Foffa:2019eeb}.

Tails present themselves with a characteristic logarithmic and hereditary nature, \emph{i.e.}, which depends of the entire history of the source rather than its state at the retarded time, corresponding to wave propagation inside the retarded light-cone. We focus our investigation in the present work to such tail logarithmic contributions to the conservative dynamics. In particular we elaborate on a result presented in \cite{Foffa:2019eeb} and give a formal presentation of simple tail contributions to the generic conservative dynamics, in the form of an action valid in principle at all PN orders.

As an application we recover the known logarithmic tail terms in the energy function of circular binaries at the 4PN and 5PN orders~\cite{BDLW10b, D10sf, LBW12, Foffa:2019eeb}, and we obtain the new results for the logarithmic tail terms at the 6PN (beyond the self-force (SF) approximation) and 7PN orders. However we know that in the latter 7PN case, which corresponds to 3PN beyond the leading 4PN logarithm, the iterated tail-of-tail-of-tail~\cite{MBF16} process is also relevant, so the complete 7PN result, derived from first principles, will have to wait an investigation of this process in the energy function. Nevertheless, by resorting to a variety of methods (traditional PN computation, EFT renormalization group flow and input from SF calculations), we manage to derive all of the logarithmic energy terms at this order. The latter result has been extended by computing the contribution of the leading $(\log)^n$ terms for all $n$, using renormalization group techniques. Note that the tail-of-tail process does not induce logarithmic terms, and contributes only to half-integer PN approximations~\cite{BFW14a, BFW14b}. 

The paper is structured as follows. In Sec.~\ref{sec:tail} we formally derive with EFT methods the tail contribution to the conservative action to all PN order; in Sec.~\ref{sec:nonlocal} we give a detailed derivation of the resulting (non-local) dynamics; in Sec.~\ref{sec:circular} we specialize to binaries on circular orbits and derive the energy function up to 7PN order; in Sec.~\ref{sec:RGtheory} we use renormalization group equations for mass and angular momentum to compute the contribution of the dominant $(\log)^n$ terms in the invariant energy of circular orbits; finally, we conclude in Sec.~\ref{sec:conclusions}. In the Appendix~\ref{sec:app_action} we give an explicit alternative proof of the action we adopted in Sec.~\ref{sec:tail} with traditional PN methods restricted at the 1PN order; and some details of lengthy computations are presented in the Appendix~\ref{sec:app_nonlocal}.

\section{Complete action for simple tail terms}
\label{sec:tail}
\subsection{Action in Fourier and time domains}

Following Ref.~\cite{Foffa:2019eeb} we explore in this paper the action for all the non-local simple tails, involving all multipolar contributions $\ell\geqslant 2$, which is part of the effective action of EFT or equivalently the Fokker action of traditional PN methods. Overlooking purely local terms, this takes the following form
\begin{equation}\label{eq:StailFourier}
S_\text{tail}^{\text{nl}} = -\sum_{\ell=2}^{+\infty} \frac{2 G^2 M}{c^{2\ell+4}}\,\int_{-\infty}^{+\infty}\frac{\dd k_0}{2\pi} \log{\left(\frac{\vert k_0\vert}{\mu}\right)} k_0^{2\ell+2}\left[ a_\ell \,\vert \tilde{I}_L(k_0)\vert^2 + \frac{b_\ell}{c^2} \,\vert \tilde{J}_L(k_0)\vert^2\right]\,,
\end{equation}
where $M$ is the ADM mass, $I_L(t)$ and $J_L(t)$ denote the mass and current type source multipole moments (with $L=i_1\cdots i_\ell$ the usual collective notation for $\ell$ independent spatial indices) with Fourier integrals
\begin{equation}\label{eq:Fourier} 
\tilde{I}_L(k_0) = \int_{-\infty}^{+\infty} \dd t \,I_L(t)\,\de^{\di k_0 t} \quad\text{and}\quad \tilde{J}_L(k_0) = \int_{-\infty}^{+\infty} \dd t \,J_L(t)\,\de^{\di k_0 t}\,.
\end{equation}
In traditional PN methods the mass and current multipoles are defined in harmonic coordinates by the metric~\eqref{eq:h1munu} below. In~\eqref{eq:StailFourier} $\mu$ is an arbitrary energy scale which in dimensional regularization relates the standard 3-dimensional Newton constant $G\equiv G_N$ to the $d$-dimensional gravitational coupling through $G^{(d)} = \mu^{3-d} G$. Finally the coefficients in~\eqref{eq:StailFourier} are exactly those which appear in the multipole expansion of the gravitational wave energy flux~\cite{Th80}, namely
\begin{equation}\label{eq:abell}
a_\ell = \frac{(\ell+1)(\ell+2)}{(\ell-1)\ell \ell!(2\ell+1)!!}\,,\qquad b_\ell = \frac{4\ell(\ell+2)}{(\ell-1)(\ell+1)!(2\ell+1)!!}\,.
\end{equation}
In our investigations below we consistently recover from the action the expression of the energy flux, see Eq.~\eqref{eq:fluxE}.

In the time domain, the non-local action~\eqref{eq:StailFourier} becomes
\begin{equation}\label{eq:Stailtimedomain}
S_\text{tail}^{\text{nl}} = \sum_{\ell=2}^{+\infty} \frac{G^2 M}{c^{2\ell+4}} \int_{-\infty}^{+\infty} \dd t \left[ a_\ell \,I_L^{(\ell+1)}(t)\,\mathcal{I}_L^{(\ell+1)}(t) + \frac{b_\ell}{c^2} J_L^{(\ell+1)}(t)\,\mathcal{J}_L^{(\ell+1)}(t) \right]\,,
\end{equation}
where the superscript $(n)$ denotes time derivatives, and we have conveniently posed
\begin{equation}\label{eq:taillogIL}
\mathcal{I}_L (t) = \int_0^{+\infty} \dd\tau
\,\log\left(\frac{\tau}{\tau_0} \right)\left[I_L^{(1)}(t-\tau) -
  I_L^{(1)}(t+\tau)\right]\,,
\end{equation}
together with the same definition for the functional $\mathcal{J}_L[J_L]$. Here the scale $\tau_0$ is related to $\mu$ by $c \tau_0\equiv\mu^{-1}\,\de^{-\gamma_\text{E}}$, with $\gamma_\text{E}$ being the Euler constant.\footnote{For a conservative dynamics, the source moments are time symmetric, $I_L(-t)=I_L(t)$, and we can check that the definition~\eqref{eq:taillogIL} is also time-symmetric, \emph{i.e.} $\mathcal{I}_L(-t)=\mathcal{I}_L(t)$.} The expression~\eqref{eq:taillogIL} is equivalent to the following form, often used in the literature, involving the so-called Hadamard finite part or partie finie (Pf) prescription in terms of the scale $\tau_0$,
\begin{equation}\label{eq:taildef}
\mathcal{I}_L (t) = 
\mathop{\text{Pf}}_{\tau_0} \int_{-\infty}^{+\infty} \frac{\dd t'}{\vert
  t-t'\vert} \,I_{L}(t')\,,
\end{equation}
in terms of which the time-domain action~\eqref{eq:Stailtimedomain} takes the elegant form
\begin{equation}\label{eq:Stail}
S_\text{tail}^{\text{nl}} = \sum_{\ell=2}^{+\infty} \frac{G^2 M}{c^{2\ell+4}}\,\mathop{\text{Pf}}_{\tau_0}\int\!\!\!\int\frac{\dd t\dd t'}{\vert t-t'\vert} \left[ a_\ell \,I_L^{(\ell+1)}(t)I_L^{(\ell+1)}(t') + \frac{b_\ell}{c^2} J_L^{(\ell+1)}(t)J_L^{(\ell+1)}(t') \right]\,.
\end{equation}

In the derivation of the 4PN equations of motion either by Hamiltonian~\cite{JaraS13, DJS14, JaraS15, DJS16}, Lagrangian~\cite{BBBFMa, BBBFMb, BBBFMc, MBBF17} or EFT~\cite{FS4PN, FMSS16, FS19, FPRS19} methods, it was proved that (i) the unphysical scale $\mu$ (or equivalently $c\tau_0=\mu^{-1}\de^{-\gamma_\text{E}}$) originally present in the tail action~\eqref{eq:StailFourier} finally disappears from the final total action; (ii) the role of the cut-off scale in the logarithmic term of the tail part of the final action is played by the distance between the two bodies, $r \equiv r_{12} = \vert\bm{y}_1-\bm{y}_2\vert$. Therefore we shall from now fix the unphysical scale $\tau_0$ in Eq.~\eqref{eq:Stail} to be $\tau_0=r/c$, where $r$ is the distance in harmonic coordinates. For a more detailed derivation of the substitution $\tau_0\to r/c$, due to the interplay between near and far zone logarithms, we refer the interested reader to Sec.~IV of~\cite{FS19}.

\subsection{Proof of the action by EFT methods}
\label{sec:proof}

Following Ref.~\cite{Foffa:2019eeb}, we show a general proof of the action~\eqref{eq:StailFourier} based on EFT methods.
\begin{figure}
\includegraphics[width=.35\linewidth]{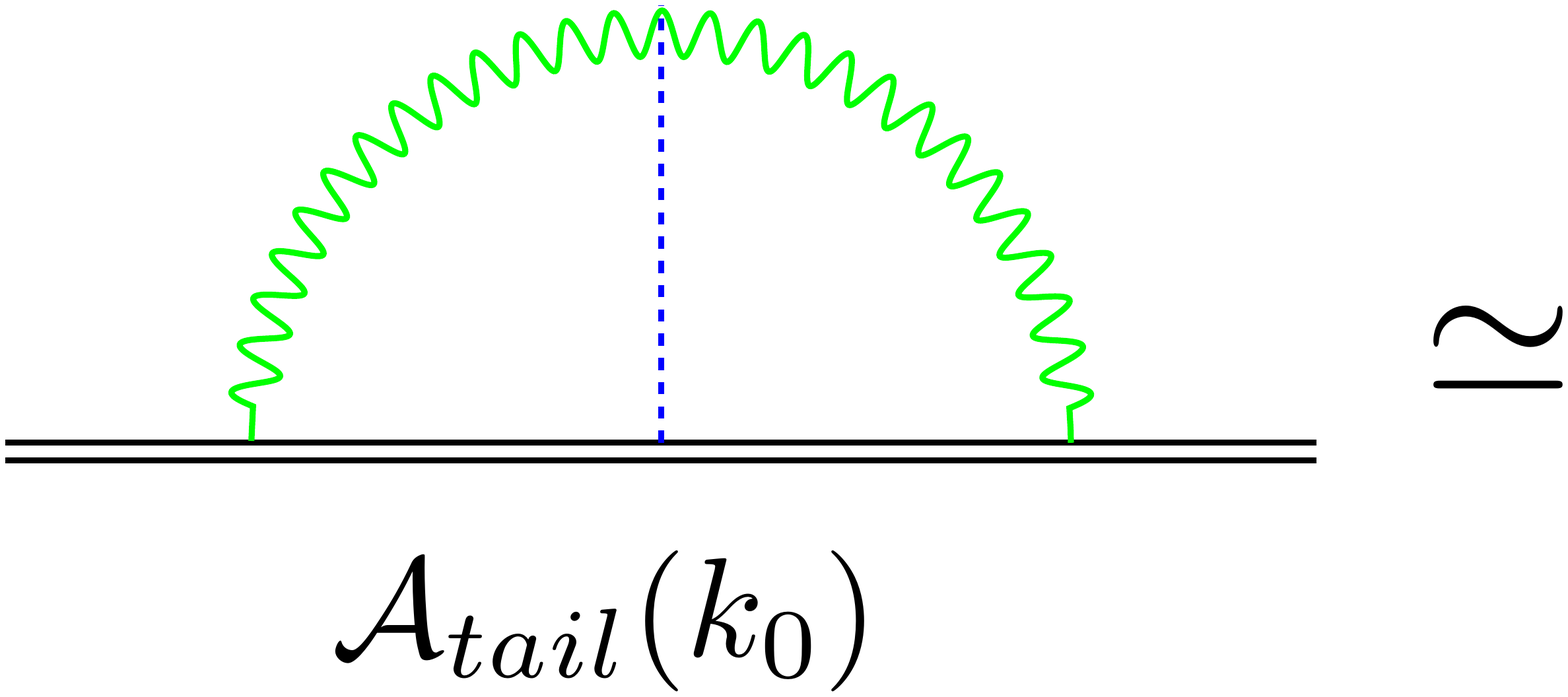}\includegraphics[width=.35\linewidth]{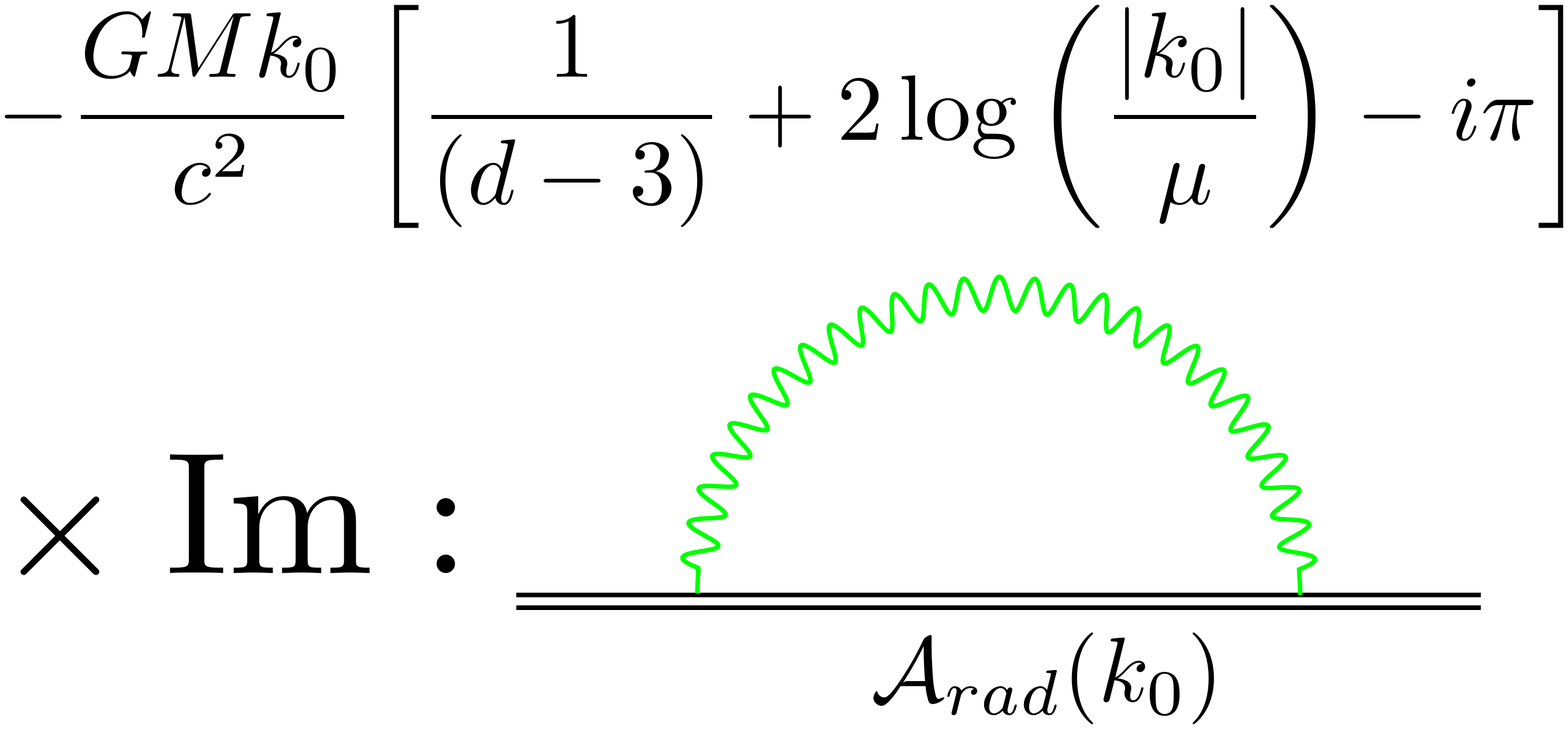}
\caption{Schematic representation of the relation between the tail self-energy amplitude (which includes also a real finite, local term understood here) and the purely imaginary self-energy diagram. The logarithmic term fully determines the non-local tail action ~\eqref{eq:StailFourier}; the imaginary part relates to the tail contribution to the energy flux, which is proportional \emph{via} a factor of $GMk_0\pi$ to the non-tail radiation flux; while the UV pole cancels the IR pole coming from near-zone conservative contributions~\cite{JaraS13, DJS14, JaraS15, DJS16, BBBFMa, BBBFMc, MBBF17, PR17, FPRS19}.} \label{fig:hereditary}
\end{figure}
Such result is a direct consequence of the more general relation represented in the Figure~\ref{fig:hereditary}, relating the singular, logarithmic and imaginary part of the amplitude of the self-energy tail process $\mathcal{A}_\text{tail}(k_0)$ to $k_0 \mathcal{A}_\text{rad}(k_0)$, the imaginary part of the latter providing the energy spectrum of the GW radiation emitted by the system. Indeed, a direct calculation of the latter shows that for every multipole $\ell$,\footnote{We use a mostly plus signature convention. To lighten the formulae we adopt the notation $\int_\K\equiv \int \frac{\dd^3k}{(2\pi)^3}$ and when propagators in momentum space are involved, the $-i0^+$ is always understood, \emph{i.e.}, $\K^2-k_0^2\to \K^2-k_0^2-i0^+$, according to the Feynman prescription to contour the propagator pole.}
\be
\label{eq:PII}
\mathcal{A}_\text{rad}(k_0)=\ds -\frac{4\pi G_N}{c^{2\ell+2}\ell !}
\int_\K&&\!\!\!\frac{\K_{(L-2)}\K_{(L'-2)}}{\K^2-k_0^2}\left[I_{ij(L-2)}(k_0) I_{kl(L'-2)}(-k_0)K_I^{ijkl}\left(k_0,\K\right)+\right.\nonumber\\
&&+\left.\frac{16}{9c^2}J_{ij(L-2)}(k_0) J_{kl(L'-2)}(-k_0)K_J^{ijkl}\left(k_0,\K\right)\right]\,,
\ee
with $K_I^{ijkl}\equiv -2k_0^4 \delta^{ik}\delta^{jl}+4k_0^2\delta^{ik}k^jk^l
-k^ik^kk^jk^l$ and $K_J^{ijkl}=-k_0^2\K^2 \delta^{ik}\delta^{jl}+\delta^{ik}k^jk^l\left(k_0^2+\K^2\right)-k^ik^kk^jk^l+k_0^2\varepsilon^{rik}\varepsilon^{wjl}k_rk_w$, where $\varepsilon^{ijk}$ denotes the usual Levi-Civita symbol.

As to the tail part, all singular, logarithmic and imaginary terms are coming from the graviton propagator pole $\K^2\simeq k_0^2$, and that around such momentum region
\be
\label{eq:poletail}
\mathcal{A}_\text{tail}(k_0)\simeq\ds -\frac{4\pi G_N^2 M}{c^{2\ell+4}\ell !}
(-k_0^2)^{d/2-1}\int_\K &&\!\!\!\frac{\K_{(L-2)}\K_{(L'-2)}}{\K^2-k_0^2}\left[I_{ij(L-2)}(k_0) I_{kl(L'-2)}(-k_0)\hat{K}_I^{ijkl}\left(k_0,\K\right)+\right.\nonumber\\
&&\left.+\frac{16}{9c^2}J_{ij(L-2)}(k_0) J_{kl(L'-2)}(-k_0)\hat{K}_J^{ijkl}\left(k_0,\K\right)\right]f\left(\frac{\K^2}{k_0^2}\right)\,,
\ee
with $\hat{K}_I^{ijkl}\equiv-2k_0^4 \delta^{ik}\delta^{jl}+4\K^2\delta^{ik}k^jk^l
-k^ik^kk^jk^l$, $\hat{K}_J^{ijkl}\equiv[-k_0^4\K^2 \delta^{ik}\delta^{jl}+\delta^{ik}k^jk^l\left(k_0^4+\K^4\right)-k^ik^kk^jk^l \K^2+k_0^4\varepsilon^{rik}\varepsilon^{wjl}k_rk_w]/k_0^2$, so that
$\hat{K}_{I,J}^{ijkl}=K_{I,J}^{ijkl}$ on the propagator pole. 

The universal dimensionless function $f\left(\frac{\K^2}{k_0^2}\right)$ is given by
\be
\ds f\left(\frac{\K^2}{k_0^2}\right)=4 \pi  (-k_0^2-i0^+)^{2-d/2}
\int_{\Q}\frac1{\Q^2\left[(\K+\Q)^2-k_0^2\right]}\,,
\ee
and its direct evaluation in the limit $\frac{\K^2}{k_0^2}\rightarrow 1$ gives the result reported in the Figure~\ref{fig:hereditary}.

In the Appendix~\ref{sec:app_action} we will provide an alternative proof of the log terms in the action~\eqref{eq:Stail} using traditional PN methods, and valid, for every $\ell$, at next-to-leading 1PN order for mass multipoles and at leading order for current ones, using the explicit expression of the mass multipole moments at the 1PN order. This confirms that the moments $I_L$ and $J_L$ are indeed identical to the PN source multipole moments of traditional PN methods.

\section{Derivation of the dynamics for the log-tail contributions}
\label{sec:nonlocal}
\subsection{Equations of motion and conserved Noetherian energy}

In the action~\eqref{eq:Stail} the multipole moments $I_L$ and $J_L$ (and also the mass $M$) are functionals of the particles' positions, velocities, accelerations, \emph{etc}. We know that the action can be order reduced by means of the equations of motion, and that the variables can be expressed in the center of mass (CM) frame, so that the action~\eqref{eq:Stail} is an ordinary action depending on the relative separation $x^i=y^i_1-y^i_2$ of the two particles, and their relative velocity $v^i=\dd x^i/\dd t$. And, as we said, the tail term is obtained with $\tau_0=r/c$ where $r=\vert\bm{x}\vert$ is the separation between particles. We denote by $m=m_1+m_2$ the total mass and $\nu=m_1 m_2/m^2$ the symmetric mass ratio.

We vary the action~\eqref{eq:Stail} with respect to the particles' relative variables, taking into account the non-local structure of the action (see~\cite{BBBFMa} for more details). Considering the multipole moments as functionals of the independent variables $x^i$ and $v^i$ we get the direct contribution of tails in the equations of motion for the acceleration $a^i=\dd v^i/\dd t$ as\footnote{We ignore the variation of the scale $\tau_0=r/c$ since this gives an instantaneous (non-tail) term without logarithms.}
\begin{align}\label{eq:acctail}
\Delta a^i_\text{tail} &= \frac{1}{m\nu}\sum_{\ell=2}^{+\infty} \frac{G^2 a_\ell}{c^{2\ell+4}} \left\{ \frac{\partial M}{\partial x^i} I_L^{(\ell+1)}\,\mathcal{I}_L^{(\ell+1)} - \frac{\dd}{\dd t}\left(\frac{\partial M}{\partial v^i} I_L^{(\ell+1)}\,\mathcal{I}_L^{(\ell+1)}\right) - 2M(-)^{\ell}\left[ \frac{\partial I_L}{\partial x^i} \,\mathcal{I}_L^{(2\ell+2)} - \frac{\dd}{\dd t}\left(\frac{\partial I_L}{\partial v^i}\,\mathcal{I}_L^{(2\ell+2)}\right)\right]\right\}\nonumber\\
&+ \text{identical contribution for the current moments, \textit{i.e.} with $(I_L, \mathcal{I}_L, a_\ell)\longrightarrow (J_L, \mathcal{J}_L, c^{-2} b_\ell)$}\,.
\end{align}
Furthermore, at high PN order there will be also other tail contributions (not detailed here) coming from the replacements of accelerations in lower order terms of the final equations of motion, as well as coming later from the reduction to quasi circular orbits. From the tail contribution~\eqref{eq:acctail} in the acceleration we obtain the corresponding tail contribution in the conserved energy as (see~\cite{BBBFMb} for details)
\begin{align}\label{eq:DeltaE}
\Delta E_\text{tail} &= \sum_{\ell=2}^{+\infty} \frac{G^2 a_\ell}{c^{2\ell+4}}\biggl\{v^i\frac{\partial M}{\partial v^i} I_L^{(\ell+1)}\mathcal{I}_L^{(\ell+1)} + M \biggl[ 2 \sum_{p=1}^{\ell} (-)^p I_L^{(\ell+1-p)} \mathcal{I}_L^{(\ell+1+p)} + I_L^{(\ell+1)} \mathcal{I}_L^{(\ell+1)} - 2(-)^\ell v^i\frac{\partial I_L}{\partial v^i}\,\mathcal{I}_L^{(2\ell+2)} + \delta H_\ell\biggr] \biggr\}\nonumber\\
&+ \text{identical contribution for the current moments, \textit{i.e.} with $(I_L, \mathcal{I}_L, a_\ell)\longrightarrow (J_L, \mathcal{J}_L, c^{-2} b_\ell)$}\,.
\end{align}
The terms are easily derived, except the last one $\delta H_\ell$ which represents a non trivial correction to be added in the case of the non-local dynamics, for which the Noetherian conserved energy $E$ actually differs from the value of the non-local Hamiltonian $H$ computed on shell~\cite{BBBFMb}. This term satisfies
\begin{equation}\label{eq:deltaHdef}
\frac{\dd \delta H_\ell}{\dd t} = I_L^{(\ell+1)} \mathcal{I}_L^{(\ell+2)} - I_L^{(\ell+2)} \mathcal{I}_L^{(\ell+1)}\,,
\end{equation}
which is the generalisation of Eq.~(3.8) of~\cite{BBBFMb} in the case of generic multipole moments.
Interestingly, in analogy to similar relations discussed in Refs.~\cite{Blanchet:1995fr, Goldberger:2009qd, Foffa:2019eeb}, note that the time average over one period of this term is directly related to the total GW energy flux,
\begin{equation}\label{eq:average}
\langle\delta H\rangle = - \frac{2G M}{c^3}\,\mathcal{F}_\text{GW}\,,
\end{equation}
where we have naturally denoted $\delta H = \sum \frac{G^2Ma_\ell}{c^{2\ell+4}}\delta H_\ell +$ identical contribution for the current moments, and where the total (averaged) energy flux associated with the source moments $I_L$ and $J_L$ reads
\begin{equation}\label{eq:fluxE}
\mathcal{F}_\text{GW} = \sum_{\ell=2}^{+\infty} \frac{G}{c^{2\ell+1}} \left[ a_\ell \,I_L^{(\ell+1)}\,I_L^{(\ell+1)} + \frac{b_\ell}{c^2} J_L^{(\ell+1)}\,J_L^{(\ell+1)} \right]\,.
\end{equation}
This result confirms the soundness of the general action~\eqref{eq:Stail} with the general coefficients~\eqref{eq:abell}. But contrary to the other contributions in~\eqref{eq:DeltaE}, the term $\delta H_\ell$ will not contribute to the logarithmic part of the conserved energy, when dealing with quasi-circular orbits. Thus we present its computation, and justify the time averaged formula~\eqref{eq:average}, in the Appendix~\ref{sec:app_nonlocal}.

\subsection{Case of quasi-circular orbits}

The previous investigation was based on the non-local action~\eqref{eq:Stail}, but we now specialize it to quasi-circular orbits. In the case of quasi circular orbits (in the adiabatic approximation) the formulas simplify drastically and in particular the tail integrals~\eqref{eq:taildef} become local (see \emph{e.g.}~\cite{BBBFMa}):
\begin{equation}\label{eq:calIcirc}
\mathcal{I}_L = -2 I_{L}\biggl[\log\left(\frac{r\,\omega}{c}\right) + \gamma_\text{E} \biggr]\,,\qquad\mathcal{J}_L = -2 J_{L}\biggl[\log\left(\frac{r\,\omega}{c}\right) + \gamma_\text{E} \biggr]\,,
\end{equation}
where $\omega$ denotes the orbital frequency of the binary and $\gamma_\text{E}$ denotes the Euler constant. For circular orbits we have $\log(\frac{r\,\omega}{c}) = \frac{1}{2} \log\gamma$ plus some non-logarithmic contributions that we neglect, where $\gamma=G m/(r c^2)$ is the usual PN parameter in harmonic coordinates, and we can ignore the Euler constant. Hence we obtain the purely logarithmic contributions in the tail acceleration~\eqref{eq:acctail} for circular orbits as
\begin{align}\label{eq:acctailcirc}
\Delta a^i_\text{tail} =& -\frac{2}{m\nu}\sum_{\ell=2}^{+\infty} \frac{G^2 a_\ell}{c^{2\ell+4}}\,\left[ \frac{\partial M}{\partial x^i} \left(I_L^{(\ell+1)}\right)^2 - \frac{\dd}{\dd t}\left(\frac{\partial M}{\partial v^i} \left(I_L^{(\ell+1)}\right)^2\right)\right]\log\left(\frac{r\,\omega}{c}\right)\nonumber\\
&-\frac{4}{m\nu}\sum_{\ell=2}^{+\infty} \frac{G^2 M a_\ell(-)^{\ell+1}}{c^{2\ell+4}}\,\left[ \frac{\partial I_L}{\partial x^i} \,I_L^{(2\ell+2)} - \frac{\dd}{\dd t}\left(\frac{\partial I_L}{\partial v^i}\,I_L^{(2\ell+2)}\right)\right]\log\left(\frac{r\,\omega}{c}\right)\nonumber\\
&+ \text{identical contribution for the current moments}\,.
\end{align}
As for the logarithmic contributions in the conserved energy~\eqref{eq:DeltaE} for circular orbits we find
\begin{align}\label{eq:Etailcirc}
\Delta E_\text{tail} =& -2\sum_{\ell=2}^{+\infty} \frac{G^2 a_\ell}{c^{2\ell+4}}\,v^i\frac{\partial M}{\partial v^i} \left(I_L^{(\ell+1)}\right)^2\log\left(\frac{r\,\omega}{c}\right)\nonumber\\
&-4\sum_{\ell=2}^{+\infty} \frac{G^2 M a_\ell(-)^{\ell+1}}{c^{2\ell+4}}\,\left[ \sum_{p=1}^{\ell} (-)^p I_L^{(p)} I_L^{(2\ell+2-p)} - \frac{(-)^{\ell}}{2}\left(I_L^{(\ell+1)}\right)^2 +v^i\frac{\partial I_L}{\partial v^i}\,I_L^{(2\ell+2)}\right]\log\left(\frac{r\,\omega}{c}\right)\nonumber\\
&+ \text{identical contribution for the current moments}\,.
\end{align}
Remind that the extra contribution found $\delta H_\ell$ in~\eqref{eq:DeltaE}, is explicitly given in Eq.~\eqref{eq:app_deltaHfinal} and does not contain logarithmic terms for circular orbits.

Equivalently, one could observe that the following local action (where $v/c=\gamma^{1/2}$ for circular orbits)
\begin{equation}\label{eq:Stailcirc}
S_\text{tail}^\text{circ} = -\sum_{\ell=2}^{+\infty} \frac{2 G^2 M}{c^{2\ell+4}}\,\int\dd t \log{\left(\frac{v}{c}\right)}\left[ a_\ell \,\left( I_L^{(\ell+1)}(t)\right)^2+ \frac{b_\ell}{c^2} \,\left(J_L^{(\ell+1)}(t)\right)^2\right]\,,
\end{equation}
leads to the same logarithmic contributions for gauge invariant quantities as the non-local one, in the case of quasi-circular motion (in the adiabatic approximation). This is so because on such particular solution of the equations of motion, one has
\be
\varphi_\ell(\tau)\equiv I^{(\ell+1)}_L(t)I^{(\ell+1)}_L(t+\tau)=\sum_{\ell'\leqslant \ell+1} \varphi_{\ell'}(0) \cos{\left(\ell'\omega\tau\right)}\,,
\ee
where $\ell'$ has the same (opposite) parity than $\ell$ for mass (current) multipoles, and $\varphi_{\ell'}$ is the part of $\varphi_\ell$ characterized by the frequency $\ell'\omega$. Plugging this relation into Eq.~\eqref{eq:Stail}\footnote{This is justified as the circular ansatz is a solution of the equations of motion, its replacement into the action is equivalent to a coordinate shift, which does not change gauge invariants quantities.} and observing that
\be
\mathop{\text{Pf}}_{\tau_0}\int \frac{\dd\tau}{\vert\tau\vert}\cos(\omega \tau) = -2\log(\omega\tau_0) + \text{non-log terms}\,,
\ee
one obtains the result Eq.~(\ref{eq:Stailcirc}) after discarding terms involving $\ell'$ from the logarithmic part, as $\log(\ell'\omega\tau_0)=\log(\omega\tau_0)+\log{\ell'}$, and combining the remaining overall $\log(\omega\tau_0)$ with the $\log(\frac{r}{c\tau_0})$ coming from the near zone contributions~\cite{FPRS19} (which is mathematically equivalent to set $\tau_0=\frac rc$). In summary, according to this simplified version of the argument presented in~\cite{DJS16}, Eq.~\eqref{eq:Stailcirc} is equivalent to the original non-local action, as long as it is employed only to derive gauge invariant logarithmic contributions in the quasi-circular regime.

\section{Logarithmic and half-integer contributions in the circular energy}
\label{sec:circular}
\subsection{Simple tail terms}
\label{subs:tail}

As an application, we proceed to compute logarithmic contributions to the circular energy, as a function of the orbital frequency, or of the equivalent dimensionless variable $x\equiv (G m\omega/c^3)^{2/3}$. Up to now, such contributions have been computed up to 5PN order~\cite{BDLW10b, D10sf, LBW12}, while at 6PN order only the leading term in the symmetric mass ratio $\nu\equiv\frac{m_1 m_2}{m^2}$ is known~\cite{Bini:2013rfa}. The present work, in particular the material displayed in Sec.~\ref{sec:nonlocal}, provides the ideal tool to compute all the logarithmic terms coming from simple tails, the only limitation being the knowledge of the multipole moments of the binary constituents, at the desired PN order. Given that the current knowledge is limited to 3PN for the mass quadrupole moment~\cite{BIJ02, BI04mult, FBI15}, we are presently able to provide such terms up to 7PN order.

From the EFT side, logarithmic contributions to the energy are associated with UV divergent diagrams in the far zone (or equivalently to IR divergences in the near zone), their divergence being compensated by an opposite IR divergence from the near zone, the logarithmic terms from near and far zone combine to give a $\log(k_0 r)$ term~\cite{FPRS19}. Self-energy diagrams with bulk interactions producing tail terms give rise to divergent terms when integration over past time extends to present time [\emph{i.e.} $t'\to t$ in Eq.~\eqref{eq:Stail}]. As explicitly shown in~\cite{Foffa:2019eeb}, this does not happen with leading order memory diagrams, which give instantaneous contributions to the self energy and to ``failed'' tail diagrams involving an angular momentum instead of the mass $M$ at the insertion of the blue dashed line onto the source in Fig.~\ref{fig:hereditary}. However additional sub-leading logarithmic terms are expected from tail-of-tail and mixed tail-memory processes.

Similarly, in traditional PN methods, all the logarithms are generated by tails propagating in the far zone, as well as iterated tails-of-tails and sub-leading tail-memory couplings.
Notice that in intermediate steps of the calculation there are other logarithms which appear, but these are pure gauge and cancel out in gauge invariant quantities. This is the case of the logarithms at the 3PN order in the equations of motion in harmonic coordinates, which disappear from the invariant circular energy and angular momentum~\cite{BlanchetLR}. In the present paper we assumed (rather than explicitly checked) that these gauge logarithms properly cancel up to 7PN order.

We have done the computation of the 7PN simple tails, in which we have to order-reduce the derivatives of the multipole moments by means of the equation of motion, using either the equations of motion obtained from the non-local formulation, or the equivalent (valid for circular orbits only) local Lagrangian given by Eq.~\eqref{eq:Stailcirc}. 
Both the variations of the detailed computation discussed above converge to the following simple-tail contributions to the logarithmic part of $E(x)$ up to 7PN order:\footnote{We set $c=1$ in Secs.~\ref{sec:circular} and~\ref{sec:RGtheory}.}
\begin{align}\label{eq:Eofxtail}
E_\text{simple-tail}^\text{log} &= -\frac{m\nu^2}{2} x^5 \log{x}\left[\frac{448}{15}+\left(-\frac{4988}{35}-\frac{656}{5}\nu\right)x+\left(-\frac{1967284}{8505}+\frac{914782}{945}\nu+\frac{32384}{135}\nu^2\right)x^2\right.\nonumber\\
&\qquad\qquad\qquad\qquad+ \left(\frac{16785520373}{2338875}-\frac{1424384}{1575}\log\left(\frac {r}{r_0}\right)+\left(\frac{2132}{45}\pi^2 -\frac{41161601}{51030}\right)\nu\right. \nonumber\\
&\qquad\qquad\qquad\qquad\left.\left.-\frac{13476541}{5670}\nu^2-\frac{289666}{1215}\nu^3\right)x^3 + \mathcal{O}\left(x^4\right) \right]\,,
\end{align}
where $r_0$ is the UV regulator which appears in the expression of the 3PN mass quadrupole moment, see \emph{e.g.}~\cite{BlanchetLR}. One expects that in the EFT computation of the multipole moments (not yet available at the 3PN order needed here), the constant $r_0$ should be related to the dimensional regularization constant $\mu$. Such constant should be cancelled by the addition of tail-of-tail-of-tail contributions, see also subsection~\ref{subs:tail3} below.

The 4PN and 5PN coefficients, as well as the leading order 6PN coefficient in the test mass limit $\nu\to 0$, confirm previous findings in the literature~\cite{BDLW10b, D10sf, LBW12, Bini:2013rfa}, while all the remaining ones are derived here for the first time.

\subsection{Tail-of-tail terms}
\label{subs:tail2}

Besides the simple tail terms, iterated tails do also contribute. The simplest of them are the so-called tail-of-tails [or (tail)$^2$] which arise one and a half order (1.5PN) beyond the simple tails, which means 5.5PN in the conserved energy corresponding to 3PN in the asymptotic waveform~\cite{B98tail,Goldberger:2009qd}. Note that the tail-of-tails arise at half-integer PN approximations in the conservative dynamics, which is possible because of the non-locality involved.\footnote{The fact that the conservative dynamics contains half-integer PN approximations starting from the 5.5PN order has been discovered in high-precison numerical SF calculations~\cite{SFW14}.} 

The (tail)$^2$ will not bring any logarithmic dependence in the conserved energy, at least up to 7.5PN order (see~\cite{BFW14a, BFW14b} for discussion), as they involve in the near-zone metric (say, the 00 component of the metric) an even number of time derivatives of the quadrupole moment. In this case, using the contractions of the moment with the field point, it is straightforward to see that the tail integral, when reduced to circular orbits, does not produce a logarithm but that a factor $\pi$ is generated instead. By contrast the simple tails at the 4PN order and the tail-of-tail-of-tails at 7PN order which will be discussed in Sec.~\ref{subs:tail3} [and in fact, any iterated $(\text{tail})^p$ with $p=2n-1$ odd, arising at order $(3n+1)$PN, see Sec.~\ref{sec:RGtheory}] involve an odd number of time derivatives of the quadrupole moment, and the tail integrals produce logarithmic terms.

The (tail)$^2$ have been computed using traditional PN methods at the leading order when $\nu\to 0$, \emph{i.e.} in the gravitational self-force (SF) limit, in the redshift variable~\cite{Det08}. Nevertheless, it is possible to deduce the (tail)$^2$ in the energy function from the corresponding result in the redshift variable thanks to the first law of binary point-like particle mechanics. As far as we know, this has not been done yet, thus we present the derivation in the present section.

In the non-spinning case, the first law of binary point-like particle mechanics~\cite{FUS02, LBW12} relates the variation of the total ADM mass $M$ and the total angular momentum $J$ to the variation of the individual masses $m_1$ and $m_2$ as
\begin{equation}\label{eq:FirstLaw}
\delta M - \omega \,\delta J = z_1 \,\delta m_1+z_2\, \delta m_2\,,
\end{equation}
where $\omega$ denotes the circular orbital frequency and $z_1$ and $z_2$ are the gravitational redshift variables. In the SF limit, the expression for those variables is known analytically from usual PN methods using dimensional regularization up to 3PN order~\cite{BDLW10a} (the log coefficients at 4PN and 5PN being added in~\cite{BDLW10b, LBW12}), from numerical SF methods up to 22PN~\cite{KOW15}, and from analytical ones up to 9.5PN~\cite{BiniDamour15}. It reads
\begin{equation}\label{eq:z1}
z_1 = \sqrt{1-3x} + \nu \,z_\text{SF}(x) + \mathcal{O}(\nu^2)\,,
\end{equation}
where $\sqrt{1-3x}$ is the Schwarzschild redshift in the test-mass limit. The tail-of-tails in the SF part of the redshift variable are known analytically up to the 2PN relative order, and read~\cite{BFW14a,BFW14b}\footnote{Note that the ``redshift variable'' used in those references is $u_1^T = 1/z_1$ and is expressed in terms of $y = x(1+m_1/m_2)^{-2/3}$, thus the different numerical coefficients; here $m_1\ll m_2$ is the smaller mass orbiting the larger one $m_2$, say the black hole.}
\begin{equation}\label{eq:zSFtails2}
z_\text{SF}^\text{(tail)$^2$} = \pi\,x^{13/2}\left[ \frac{13696}{525}-\frac{368693}{3675}\,x-\frac{361209292}{3274425}\,x^2+ \mathcal{O}\left(x^3\right)\right]\,.
\end{equation}
Note that these terms represent the full contributions in the redshift variable to these orders, and they are in agreement with modern analytic SF computations of the redshift up to high PN order~\cite{KOW15}.

Integrating the first law~\eqref{eq:FirstLaw}, it is possible to express the conserved energy of the particle orbiting around a big black hole in terms of the redshift variable of that particle as~\cite{LBB12}
\begin{equation}
\frac{E}{m\nu} = \frac{1-2x}{\sqrt{1-3x}}-1 + \nu\,\mathcal{E}_\text{SF}(x) + \mathcal{O}\left(\nu^2\right)\,,
\end{equation}
where the interesting contribution reads
\begin{equation}
\mathcal{E}_\text{SF}(x) = \frac{1}{2}\,z_\text{SF}(x)-\frac{x}{3}\,z_\text{SF}'(x)+\sqrt{1-3x}-1+\frac{x}{6}\frac{7-24x}{(1-3x)^{3/2}}\,.
\end{equation}
Plugging the redshift~\eqref{eq:zSFtails2} in this expression, the energy contribution of the tail-of-tails at leading order in $\nu$ reads
\begin{equation}\label{eq:Extail2}
E_\text{(tail)$^2$}(x) = -\frac{m \nu\,x}{2} \left\lbrace
\nu\,\pi\,x^{11/2}\biggl[\frac{27392}{315}-\frac{1474772}{3675}\,x-\frac{722418584}{1403325}\,x^2
+ \mathcal{O}\left(x^3\right)\biggr] + \mathcal{O}\left(\nu^2\right)\right\rbrace\,.
\end{equation}
From the 6.5PN order we also expect the coupling between the simple tail and the memory effect to contribute to the conservative dynamics. However such ``tail-of-memory'' effect does not enter the formulas~\eqref{eq:zSFtails2} and~\eqref{eq:Extail2}, as it only affects terms of higher-order in $\nu$.

\subsection{7PN logs and tail-of-tail-of-tail terms}
\label{subs:tail3}

The simple tail-logs contributions displayed in Eq.~\eqref{eq:Eofxtail} are the only logarithmic terms contributing to the observable $E(x)$ up to 6PN order, while at 7PN one should account also for the leading order tail-of-tail-of-tail or (tail)$^3$ terms, which are expected to cancel out the residual UV regulator $r_0$ from Eq.~\eqref{eq:Eofxtail}.

The leading order (tail)$^3$ contribution to the energy, which contains both $\log$ and $(\log)^2$ terms, is purely quadratic in the mass ratio $\nu$ (because the quadrupole is linear in $\nu$ at leading order). This means that while the $\nu$-dependent 7PN terms in the square bracket of Eq.~\eqref{eq:Eofxtail} account for the total logarithmic contributions, this is not true for $\nu$-independent ones, so that we can write
\begin{align}\label{eq:Eofx7PNlogs}
E_\text{7PN}^\text{log}(x) &= -\frac{m\nu^2}{2} x^8 \log{x}\left[c_{7\text{PN}}^{\log^2}\log {x}+c_{7\text{PN}}^{\log}+\left(\frac{2132}{45}\pi^2 -\frac{41161601}{51030}\right)\nu-\frac{13476541}{5670}\nu^2-\frac{289666}{1215}\nu^3\right]\,.
\end{align}
The coefficient $c_{7PN}^{\text{log}^2}$ will be computed from first principles in
the next section: for the moment we just notice that both $c_{7PN}^{\log}$ and
$c_{7PN}^{\log^2}$ can be derived from the SF redshift results of \cite{HKO15};
proceeding along the same lines discussed above we find
\begin{equation}\label{eq:c7PN}
c_{7\text{PN}}^{\log} =  \frac{85229654387}{16372125} - \frac{1424384}{1575}\bigl(\gamma_\text{E}+\log 4\bigr)\,,\qquad c_{7\text{PN}}^{\log^2} =  -\frac{356096}{1575}\,,
\end{equation}
and this allows us to predict the leading order (tail)$^3$ contribution as the difference between Eq.~\eqref{eq:Eofx7PNlogs} and the 7PN part of Eq.~\eqref{eq:Eofxtail}, namely
\begin{align}\label{eq:Eofxtail3}
E_{(\text{tail})^3}^\text{log}(x)&= -\frac{m\nu^2}{2} x^8 \log{x}\left\{-\frac{356096}{1575}\log {x}-\frac{108649792}{55125} +\frac{1424384}{1575}\left[\log\left(\frac {r}{r_0}\right)- \gamma_\text{E}-\log 4\right]+ \mathcal{O}\left(x\right)\right\}\,.
\end{align}
Note that the coefficient of $\gamma_\text{E}$ in~\eqref{eq:Eofx7PNlogs} is exactly matching the one of $\ln(r/r_0)$ in Eq.~\eqref{eq:Eofxtail}, thus giving an indication that the cancellation of the UV scale by the tail-of-tail-of-tails will be straightforward.

\section{Leading logarithms from renormalization group theory}
\label{sec:RGtheory}
\subsection{Renormalization Group equations}
\label{subs:logn}

We are going to derive the $(\log)^2$ term in Eqs.~\eqref{eq:Eofx7PNlogs}--\eqref{eq:Eofxtail3} above, together with the leading $(\log)^n$ terms for all $n$, using the renormalization group (RG) equations. In general the $(\log)^n$ terms appear at each $(3n+1)$PN order due to multiple tail interactions with two powers of the (leading order) quadrupole moment.

We start from the RG equation~(19) of Ref.~\cite{Goldberger:2012kf} for the total mass-energy $M$, that we copy here:
\begin{equation}\label{eq:RGM}
	\frac{\dd\log{M(\mu)}}{\dd\log{\mu}} = -\frac{2 G^2}{5}\left[2I_{ij}^{(1)}I_{ij}^{(5)}-2I_{ij}^{(2)}I_{ij}^{(4)}+I_{ij}^{(3)}I_{ij}^{(3)}\right]\,,
\end{equation}
where $\mu$ is the renormalization scale, and both the (Bondi\footnote{At the order we are working, and restricting to the conservative dynamics, the Bondi mass can however be traded for the ADM mass.}) mass $M$ and quadrupole moment $I_{ij}$ are defined at the scale $\mu$. This equation was originally derived from the tail correction to an external gravitational mode coupling to a source endowed with a quadrupole moment and agrees with Eq.~(4.6) of Ref.~\cite{BBFM17}, with the obvious replacement $r_{12}\rightarrow\mu$ since we are only interested in the running with the scale $\mu$. Furthermore we shall need the similar equation for the angular momentum $J^i$, which is the consequence of Eq.~(4.15a) of~\cite{BBFM17} and reads
\begin{equation}\label{eq:RGJ}
\frac{\dd J^i(\mu)}{\dd \log{\mu}} = -\frac{8G^2M}{5}\varepsilon^{ijk}\left[ I_{jl}I_{kl}^{(5)}-I_{jl}^{(1)}I_{kl}^{(4)}+I_{jl}^{(2)}I_{kl}^{(3)}\right]\,.
\end{equation}
Next, it is crucial to observe that the quadrupole moment itself undergoes a logarithmic renormalization under the RG flow, which is computable from the singularities in the scattering of gravitational radiation off the static gravitational potential at $(GM\omega)^2$ order. The quadrupole moment $I_{ij}$ at the scale $\mu$ is related to the same quantity defined at the scale $\mu_0$, and reads in the Fourier domain (with $\Omega$ the Fourier frequency), as reported by the equation~(21) of Ref.~\cite{Goldberger:2012kf} which we repeat here:
\begin{equation}
\tilde{I}_{ij}(\Omega,\mu) = \bar{\mu}^{\beta_I(G M \Omega)^2}\tilde{I}_{ij}(\Omega,\mu_0)\,,
\end{equation}
where $\bar{\mu}\equiv\mu/\mu_0$ and $\beta_I=-\frac{214}{105}$ is the coefficient associated with the logarithmic renormalization of the mass quadrupole moment~\cite{Goldberger:2009qd}. The latter relation can be written in the time domain as
\begin{equation}
\label{eq:RGI}
I_{ij}(t,\mu) = \sum_{n=0}^{+\infty} \frac{1}{n!} \bigl(-\beta_I G^2 M^2 \log\bar{\mu}\bigr)^n I_{ij}^{(2n)}(t,\mu_0)\,.
\end{equation}
Short-circuiting Eqs.~\eqref{eq:RGM} and \eqref{eq:RGI} one can derive the ADM mass renormalization group flow equation, which reads
\begin{equation}
\frac{\dd \log{M(\mu)}}{\dd \log{\bar{\mu}}} = -\frac{2G^2}{5}\sum_{k, p\geqslant 0}\frac{\bigl(-\beta_I G^2M^2\log \bar{\mu}\bigr)^{k+p}}{k!p!}\left[2I_{ij}^{(2k+1)}I_{ij}^{(2p+5)}-2I_{ij}^{(2k+2)}I_{ij}^{(2p+4)}+I_{ij}^{(2k+3)}I_{ij}^{(2p+3)}\right](t,\mu_0)\,,
\end{equation}
and by integrating we obtain the following solution:
\begin{align}
&\log\frac{M(\mu)}{M} = -\frac{2G^2}{5}\sum_{k, p\geqslant 0}\frac{\bigl(-\beta_I G^2 M^2\bigr)^{k+p}(\log \bar{\mu})^{k+p+1}}{k!p!(k+p+1)}\left[2I_{ij}^{(2k+1)}I_{ij}^{(2p+5)}-2I_{ij}^{(2k+2)}I_{ij}^{(2p+4)}+I_{ij}^{(2k+3)}I_{ij}^{(2p+3)}\right]\\
&\qquad =\frac{2}{5\beta_I M^2}\sum_{n\geqslant 0}\frac{(-\beta_I G^2 M^2 \log \bar{\mu})^{n}}{n!}\sum_{p=0}^{n-1}{{n-1}\choose{p}}\left[2I_{ij}^{(2n-2p-1)}I_{ij}^{(2p+5)}-2I_{ij}^{(2n-2p)}I_{ij}^{(2p+4)}+I_{ij}^{(2n-2p+1)}I_{ij}^{(2p+3)}\right]\nonumber\,,
\end{align}
in which the quadrupole moment $I_{ij}$ in the right-hand side and $M$ are defined at the scale $\mu_0$. Finally we can average the previous result over an orbital period for general orbits, and we approximate $\log\frac{M(\mu)}{M} \simeq \frac{M(\mu)}{M}-1$ (\emph{i.e.} discarding all terms higher than quadratic in the quadrupole moment), to obtain~\cite{Goldberger:2012kf}
\begin{equation}\label{eq:avM}
\left\langle\frac{M(\mu)}{M}\right\rangle = 1 - G^2\sum_{n=1}^{+\infty}\frac{(2\log \bar{\mu})^{n}}{n!}\bigl(\beta_I G^2 M^2\bigr)^{n-1}\langle I_{ij}^{(n+2)}I_{ij}^{(n+2)}\rangle\,,
\end{equation}
where the brackets $\langle\rangle$ denote the time-average. Exactly the same procedure applied to the angular momentum, \emph{i.e.} starting from Eq.~\eqref{eq:RGJ}, gives similarly
\begin{equation}\label{eq:avJ}
\langle J^i(\mu)\rangle = \langle J^i(\mu_0)\rangle - \frac{12G^2 M}{5}\varepsilon^{ijk}\sum_{n = 1}^{+\infty} \frac{(2\log \bar{\mu})^n}{n!} \bigl(\beta_I G^2 M^2\bigr)^{n-1} \langle I^{(n+1)}_{jl}I^{(n+2)}_{kl}\rangle\,.
\end{equation}

\subsection{Leading $\log^n$ terms at any PN order for circular orbits}

We shall now work out the consequences of the two results~\eqref{eq:avM}--\eqref{eq:avJ} for the case of circular orbits. In this case there is no need of averaging and we no longer mention the time-average process $\langle\rangle$. To reduce the latter equations to the case of circular orbits, we substitute the leading order contribution for the quadrupole moment of circular motion as given by
\be\label{eq:circ}
I_{ij}^{(n+2)}I_{ij}^{(n+2)} = 2^{2n+3}m^2 \nu^2\frac{(G m)^{n+2}}{r^{3n+2}}\,,\qquad \varepsilon^{ijk}I_{jl}^{(n+1)}I_{kl}^{(n+2)} = 2^{2n+2}m^2 \nu^2\frac{(G m)^{n+1}}{r^{3n-1}}\,\omega\,\hat J^i\,,
\ee
with $\hat J^i$ denoting the constant unit normal to the orbital plane such that $\varepsilon^{ijk} x^j v^k = r^2\omega\,\hat J^i$, where $r$ is the separation distance and $\omega$ the orbital frequency. Working out the relations~\eqref{eq:avM}--\eqref{eq:avJ} for the leading tail logarithms for circular orbits, we can approximate $M$ by $m$ in sub-leading terms, and we add also the Newtonian result. We thus obtain for $E=M-m$ and the norm $J$ (\emph{i.e.} $J^i=J\,\hat J^i$):
\begin{subequations}\label{eq:MJcirc}
	\begin{align}
	E &= \frac{1}{2} m\nu \,r^2\omega^2-\frac{G m^2\nu}{r} - 8m \nu^2 \frac{\gamma^2}{\beta_I} \sum_{n = 1}^{+\infty} \frac{1}{n!}\bigl(8\beta_I\gamma^3\log v\bigr)^n\,,\\
	J &= m\nu \,r^2\omega - \frac{48}{5} G^2 m^3 \nu^2 \frac{\omega}{\beta_I\gamma} \sum_{n = 1}^{+\infty} \frac{1}{n!}\bigl(8\beta_I\gamma^3\log v\bigr)^n\,,
	\end{align}
\end{subequations}
where we have set for convenience $\gamma= G m/r$. In~\eqref{eq:MJcirc} we have fixed the scale ratio $\bar{\mu}$ to be the relevant one for our purpose, that is the ratio between the radiation zone scale $\mu\simeq\lambda_\text{GW}^{-1}$, where the observer is located, and the orbital scale $\mu_0\simeq r^{-1}$ at which the equations~\eqref{eq:circ} hold. 
Hence we can take $\bar{\mu}\simeq r\omega=v$, where $v$ is the orbital velocity.\footnote{In the Appendix~\ref{sec:app_action}, we compute the conservative part of the metric at a generic distance $\vert\mathbf{x}\vert$ in the near zone, which contains logarithms of the type $\log(\vert\mathbf{x}\vert\lambda^{-1}_\text{GW})$. When evaluated at the location of particles, the metric yields log-terms in the equations of motion and conserved quantities equal to $\log(r/\lambda_\text{GW})=\log v +\text{const}$, in agreement with the previous argument; see the discussion after~\eqref{eq:h2} in the Appendix~\ref{sec:app_action}.}

Furthermore we also know that for circular orbits the two invariants $E(\omega)$ and $J(\omega)$ are not independent but are linked by the ``thermodynamic'' relation
\be\label{eq:thermo}
\frac{\dd E}{\dd \omega} = \omega \frac{\dd J}{\dd \omega}\,,
\ee
which is that aspect of the first law of binary black hole mechanics~\eqref{eq:FirstLaw} for which the individual masses do not vary. 

The three equations~\eqref{eq:MJcirc} and~\eqref{eq:thermo} then permit to determine the orbital separation $r$ or equivalently $\gamma$, which is defined here in harmonic coordinates, as a function of the orbital frequency:\footnote{The crucial relation between the separation and the orbital frequency for circular orbits, as well as the angular momentum RG equation, is not discussed in Ref.~\cite{Goldberger:2012kf}. We find that
working simply with the RG equation~\eqref{eq:avM} for the mass-energy does not allow to get the correct result for the invariant circular energy $E(x)$, Eq.~\eqref{eq:Eres}, for every $n$.}
\be
\label{eq:romega}
\gamma(x) = x \biggl[ 1 +  \frac{32 \nu}{15} \sum_{n = 1}^{+\infty} \frac{3n-7}{n!}(4\beta_I)^{n-1} x^{3n+1}(\log x)^n\biggr]\,,
\ee
together with the two invariants $E(x)$ and $J(x)$, that are related to each other by Eq.~\eqref{eq:thermo}:
\begin{subequations}
  \begin{align}
    \label{eq:Eres}
    E(x) &= -\frac{m\nu \,x}{2}\biggl[ 1 +  \frac{64 \nu}{15} \sum_{n = 1}^{+\infty} \frac{6n+1}{n!}(4\beta_I)^{n-1} x^{3n+1}(\log x)^n\biggr]\,,\\
    \label{eq:Jres}
    J(x) &= \frac{m^2 \nu}{\sqrt{x}}\biggl[ 1 -  \frac{64 \nu}{15} \sum_{n = 1}^{+\infty} \frac{3n+2}{n!}(4\beta_I)^{n-1} x^{3n+1}(\log x)^n\biggr]\,.
  \end{align}
\end{subequations}
Summarizing, we have obtained the leading powers of the logarithms $(\log)^n$ in the invariant energy function for circular orbits in the following form, which can quite remarkably be explicitly resummed (recall that $\beta_I=-\frac{214}{105}$) as
\begin{align}
E_\text{leading-$(\log)^n$}&= -\frac{m\nu \,x}{2}\biggl[ \frac{64 \nu}{15} \sum_{n = 1}^{+\infty} \frac{6n+1}{n!}(4\beta_I)^{n-1} x^{3n+1}(\log x)^n\biggr]\nonumber\\
&=-\frac{8m\nu^2 x^2}{15\beta_I}\biggl[\bigl(1+24 \beta_I x^3\log{x}\bigr)x^{4 \beta_I x^3}-1\biggr]\label{eq:Eresum}\,,
\end{align}
and similarly for the angular momentum,
\begin{equation}\label{eq:Jresum}
J_\text{leading-$(\log)^n$} =-\frac{32m^2\nu^2 \sqrt{x}}{15\beta_I}\biggl[\bigl(1+6 \beta_I x^3\log{x}\bigr)x^{4 \beta_I x^3}-1\biggr]\,.
\end{equation}

For the linear and quadratic log-terms ($n=1,2$) one recovers the coefficients $c_\text{4PN}^{\log}=\frac{448}{15}$ and $c_\text{7PN}^{\log^2}=-\frac{356096}{1575}$ displayed in Eqs.~\eqref{eq:Eofxtail} and~\eqref{eq:c7PN} of the previous section, while further comparison with Ref.~\cite{KOW15} (see Appendix B and the electronic archive Ref.~[19] there) shows perfect match, \textit{via} the first law of binary dynamics, between the first terms of the infinite series \eqref{eq:Eres} and the high accurate self-force results, that is up to $n\leqslant 7$. This shows the great consistency between EFT methods which predict the RG equations~\eqref{eq:avM}--\eqref{eq:avJ}~\cite{Goldberger:2009qd, Goldberger:2012kf}, the traditional PN approach which derived the first law of binary mechanics~\cite{LBW12}, and the state-of-the-art 22PN  accurate SF calculations~\cite{KOW15}.

\section{Conclusions}
\label{sec:conclusions}
Combining EFT with traditional PN methods, this work investigated the logarithmic contributions to the two-body conservative dynamics. On the one hand, starting from an effective action for the simple tails we were able to derive the contribution of simple logarithms in the acceleration and conserved energy for general orbits up to 7PN order. On the other hand, using the renormalization group equations for the mass and angular momentum, we computed the dominant contribution to every powers of logarithms in the conserved energy. The only piece in $E(x)$ at 7PN which could not be fixed from first principles, is the coefficient linear in $\log{x}$ and quadratic in $\nu$, denoted as $c_{7PN}^{\text{log}}$ in Sec.~\ref{sec:circular}.
We know that this coefficient receives contribution from the tail-of-tail-of-tails, whose direct computation has been reserved for future work. However we have been able to determine it, therefore completing our derivation up to 7PN order, by comparison with SF calculations. The overall result for $E(x)$ can be written as 
\begin{align}
E^\text{log} &= -\frac{m\nu^2}{2} x^5 \log{x}\biggl\{\frac{448}{15}+\left(-\frac{4988}{35}-\frac{656}{5}\nu\right)x+\left(-\frac{1967284}{8505}+\frac{914782}{945}\nu+\frac{32384}{135}\nu^2\right)x^2 \nonumber\\
&\qquad\qquad\qquad\qquad+ \left[\frac{85229654387}{16372125} - \frac{1424384}{1575}\bigl(\gamma_\text{E}+\log 4\bigr)+\left(\frac{2132}{45}\pi^2 -\frac{41161601}{51030}\right)\nu-\frac{13476541}{5670}\nu^2\right.\nonumber\\
&\qquad\qquad\qquad\qquad  
\left.-\frac{289666}{1215}\nu^3-\frac{356096}{1575}\log {x}\right]x^3
+ \frac{64}{15}\sum_{n=3}^{+\infty}\frac{(6n+1)\left(4\beta_I\right)^{n-1}}{n!}\,x^{3(n-1)}(\log{x})^{n-1}
+ \cdots \biggr\}\,,
\end{align}
where the ellipsis contains next-to-leading orders in logarithms at and above 8PN order, and where $\beta_I = -214/105$, while $\gamma_\text{E}$ is the Euler constant. Although it does not contain logarithmic terms, we also have derived the contribution of the tail-of-tails at leading order in the mass ratio, up to 7.5PN order, see Eq.~\eqref{eq:Extail2}.

\acknowledgments 

F.L. would like to thank Guillaume Faye for an inspiring discussion about the emergence of the $\ln r_0$ in the radiative moments. S.F. is supported by the Fonds National Suisse and by the SwissMap NCCR; S.F. would like to thank the International Institute of Physics in Natal for hospitality and support during the final stages of this work. The work of R.S. is partially supported by Conselho Nacional de Desenvolvimento Cientifico e Tecn\'ologico.

\appendix

\section{Explicit proof of the action at the 1PN order}
\label{sec:app_action}
The action~\eqref{eq:StailFourier} for the logarithms associated with (simple) tail terms has been proven by EFT methods in the Fourier domain in Sec.~\ref{sec:proof}. In this Appendix we present a proof by the traditional PN method in the time domain. The advantage of the PN proof is that it requires the explicit expression of the multipole moments $I_L$ and $J_L$ at a given PN order, which confirms that the multipole moments in~\eqref{eq:Stail} are indeed the source moments used by PN theory, in the sense of Ref.~\cite{B98mult}. The disadvantage is that we shall be restricted to the 1PN order for general mass moments of order $\ell$, and Newtonian order for current moments of order $\ell$; furthermore we will neglect some higher non-linear terms in $G$.

The source moments $I_L$ and $J_L$ are defined from the linearized multipolar solution $h_{1}^{\mu\nu}$ of the vacuum field equations in harmonic coordinates~\cite{Th80}:\footnote{The ghotic perturbation metric $h^{\mu\nu} \equiv \sqrt{-g}g^{\mu\nu} - \eta^{\mu\nu}$ satisfies the harmonic-coordinates condition $\partial_\nu h^{\mu\nu}=0$, and is post-Minkowskian expanded as $h^{\mu\nu}=G h_1^{\mu\nu} + G^2 h_2^{\mu\nu} + \mathcal{O}(G^3)$.}
\begin{subequations}\label{eq:h1munu}
\begin{align}
h^{00}_{1} &= -\frac{4}{c^2}\sum_{\ell=0}^{+\infty}
\frac{(-)^\ell}{\ell !} \partial_L \left( \frac{1}{r} I_L
(u)\right)\,, \\
h^{0i}_{1} &= \frac{4}{c^3}\sum_{\ell=1}^{+\infty}
\frac{(-)^\ell}{\ell !}  \left\{ \partial_{L-1} \left( \frac{1}{r}
{I}^{(1)}_{iL-1}(u)\right) + \frac{\ell}{\ell+1}
\varepsilon_{ijk} \partial_{jL-1} \left( \frac{1}{r} J_{kL-1}
(u)\right)\right\}\,, \\
h^{ij}_{1} &= -\frac{4}{c^4}\sum_{\ell=2}^{+\infty}
\frac{(-)^\ell}{\ell !}  \left\{ \partial_{L-2} \left( \frac{1}{r}
{I}^{(2)}_{ijL-2}(u)\right) + \frac{2\ell}{\ell+1}
\partial_{kL-2} \left( \frac{1}{r} \varepsilon_{kl(i}
{J}^{(1)}_{j)L-2}(u)\right)\right\}\,,
\end{align}
\end{subequations}
with $u=t-r/c$. The logarithmic tail terms come from quadratic interactions between the ADM mass $M$ (equal to the monopole $M\equiv I$) and the moments $I_L$ and $J_L$ for $\ell\geqslant 2$. They obey the wave equation $\Box h_{2}^{\mu\nu} = r^{-2}Q_{2}^{\mu\nu}(\bm{n},u) + \mathcal{O}(r^{-3})$ and are generated only from the leading $r^{-2}$ piece in the quadratic source, given by 
\begin{equation}\label{eq:Q2}
Q_{2}^{\mu\nu} = \frac{4 G M}{c^4}\,\frac{\dd^2 z_{1}^{\mu\nu}}{\dd t^2} + \frac{k^\mu k^\nu}{c^2}\sigma\,.
\end{equation}
The first term generates the tails properly speaking, while the second one is associated with the stress-energy tensor of gravitational waves. Here $k^\mu=(1,\bm{n})$ is the Minkowskian outgoing null vector, and 
\begin{equation}\label{eq:sigma}
	\sigma(\bm{n}, u) = \frac{1}{2}\frac{\dd z_{1}^{\mu\nu}}{\dd t}\frac{\dd z_{1\mu\nu}}{\dd t} -\frac{1}{4}\frac{\dd z_{1\mu}^{\mu}}{\dd t}\frac{\dd z_{1\nu}^{\nu}}{\dd t}\,,
\end{equation}
is proportional to the gravitational wave flux in the direction $\bm{n}$ and at retarded time $u$, as computed with the linearized approximation~\eqref{eq:h1munu}. Note that this term is also responsible for the memory effect~\cite{BD92}. In Eqs.~\eqref{eq:Q2} and~\eqref{eq:sigma} we define the $r^{-1}$ part of the (non-static) linearized metric~\eqref{eq:h1munu} to be
\begin{subequations}\label{eq:zmunu}
\begin{align}
z_{1}^{00} &= - 4 \sum_{\ell=2}^{+\infty} \frac{n_L}{\ell!
  c^{\ell+2}} {I}^{(\ell)}_{L}(u)\,,\\ z_{1}^{0i} &= - 4
\sum_{\ell=2}^{+\infty} \frac{n_{L-1}}{\ell! c^{\ell+2}} {I}^{(\ell)}_{iL-1}(u) + 4 \sum_{\ell=2}^{+\infty}
\frac{\ell}{(\ell+1)! c^{\ell+3}}\,\varepsilon_{iab}\,n_{aL-1} {J}^{(\ell)}_{bL-1}(u)\,,\\ z_{1}^{ij} &= - 4
\sum_{\ell=2}^{+\infty} \frac{n_{L-2}}{\ell! c^{\ell+2}} {I}^{(\ell)}_{ijL-2}(u) + 8 \sum_{\ell=2}^{+\infty}
\frac{\ell}{(\ell+1)! c^{\ell+3}}\,n_{aL-2}\,\varepsilon_{ab(i} {J}^{(\ell)}_{j)bL-2}(u)\,.
\end{align}
\end{subequations}
To obtain the logarithms in the near zone (when $r\to 0$) we follow the procedure detailed in Refs.~\cite{BDLW10b, LBW12}. Namely we integrate the wave equation at quadratic order using the symmetric propagator, and regularized by a ``Finite Part'' (FP) procedure to deal with the multipole expansion which is singular at the origin $r\to 0$:
\begin{equation}\label{eq:h2}
h_{2}^{\mu\nu} = \mathop{\mathrm{FP}}_{B=0}\,\sum_{k=0}^{+\infty}\left(\frac{\partial}{c\partial t}\right)^{2k}\!\!\Delta^{-k-1}\left[\left(\frac{r}{\lambda}\right)^B\frac{1}{r^2} Q_2^{\mu\nu}(\mathbf{n},u)\right]\,.
\end{equation}
The finite part depends on the typical radiation zone scale which is the gravitational wavelength $\lambda\equiv\lambda_\text{GW}$. This has been justified in Ref.~\cite{BDLW10b} in the restrictive case of exactly circular orbits with helical Killing symmetry, where the length scale $\lambda$ is introduced in the problem by the presence of the Killing vector $K^\mu\partial_\mu = \partial_t + \omega\,\partial_\varphi$ where $\omega=2\pi c/\lambda$. In the general case, note that the scale $\lambda$ in Eq.~\eqref{eq:h2} is cancelled by the purely hereditary part of the metric, \emph{i.e.} terms involving hereditary integrals of the type $\int_0^{\infty}\dd\tau\log(c\tau/\lambda)I_L(t-\tau)$.

We substitute the explicit expression~\eqref{eq:zmunu} into the source term~\eqref{eq:Q2}, expand the retardation $u=t-r/c$ when $r\rightarrow 0$, and integrate term by term using the Eqs.~(2.9)-(2.10) of~\cite{BDLW10b}. Then we look for the poles $\propto 1/B$ and gets the logarithms after applying the finite part when $B\to 0$. As a result we find that the first term in~\eqref{eq:Q2}, associated with tails, gives the following contribution to the near-zone logarithms:
\begin{subequations}\label{eq:h2tail}
	\begin{align}
	h^{00}_{2}\Big|_\text{tail} &= 8 M c^{-5} \log\left(\frac{r}{\lambda}\right) \sum_{\ell=2}^{+\infty} \frac{(-)^\ell}{\ell!} \partial_L\!\biggl[\frac{I_L^{(1)}(t-r/c) - I_L^{(1)}(t+r/c)}{r}\biggr]\,, \\
	h^{0i}_{2}\Big|_\text{tail} &= -8Mc^{-6} \log\left(\frac{r}{\lambda}\right) \sum_{\ell=2}^{+\infty} \frac{(-)^\ell}{\ell!} \partial_{L-1}\!\biggl[\frac{I_{iL-1}^{(2)}(t-r/c) - I_{iL-1}^{(2)}(t+r/c)}{r}\biggr]\nonumber\\
	& - 8M c^{-6} \log\left(\frac{r}{\lambda}\right) \sum_{\ell=2}^{+\infty} \frac{(-)^\ell}{\ell!}\frac{\ell}{\ell+1} \varepsilon_{iab}\partial_{aL-1}\!\biggl[\frac{J_{bL-1}^{(1)}(t-r/c) - J_{bL-1}^{(1)}(t+r/c)}{r}\biggr]\,,\\
	h^{ij}_{2}\Big|_\text{tail} &= 8M c^{-7} \log\left(\frac{r}{\lambda}\right) \sum_{\ell=2}^{+\infty} \frac{(-)^\ell}{\ell!} \partial_{L-2}\!\biggl[\frac{I_{ijL-2}^{(3)}(t-r/c) - I_{ijL-2}^{(3)}(t+r/c)}{r}\biggr]\nonumber\\
	& + 8M c^{-7} \log\left(\frac{r}{\lambda}\right) \sum_{\ell=2}^{+\infty} \frac{(-)^\ell}{\ell!}\frac{2\ell}{\ell+1} \partial_{aL-2}\!\biggl[\varepsilon_{ab(i}\frac{J_{j)bL-2}^{(2)}(t-r/c) - J_{j)bL-2}^{(2)}(t+r/c)}{r}\biggr]\,,
	\end{align}
\end{subequations}
%
while the second term in Eq.~\eqref{eq:Q2}, associated with the memory effect, reads
\begin{subequations}\label{eq:h2mem}
	\begin{align}
	h^{00}_{2}\Big|_\text{mem} &= -\frac{1}{2}\log\left(\frac{r}{\lambda}\right)\sum_{\ell=0}^{+\infty} (-)^\ell c^{\ell-1} \partial_L\!\biggl[\frac{\hat{\sigma}_L^{(-\ell-1)}(t-r/c) - \hat{\sigma}_L^{(-\ell-1)}(t+r/c)}{r}\biggr]\,, \\
	h^{0i}_{2}\Big|_\text{mem} &= \frac{1}{2}\log\left(\frac{r}{\lambda}\right)\sum_{\ell=0}^{+\infty} (-)^\ell c^{\ell} \partial_{iL}\!\biggl[\frac{\hat{\sigma}_L^{(-\ell-2)}(t-r/c) - \hat{\sigma}_L^{(-\ell-2)}(t+r/c)}{r}\biggr]\,,\\
	h^{ij}_{2}\Big|_\text{mem} &= -\frac{1}{2}\log\left(\frac{r}{\lambda}\right)\sum_{\ell=0}^{+\infty} (-)^\ell c^{\ell+1} \partial_{ijL}\!\biggl[\frac{\hat{\sigma}_L^{(-\ell-3)}(t-r/c) - \hat{\sigma}_L^{(-\ell-3)}(t+r/c)}{r}\biggr]\,.
	\end{align}
\end{subequations}
We have defined $\hat{\sigma}_L$ to be the symmetric-trace-free (STF) coefficients in the multipolar decomposition of $\sigma$ defined by
\begin{equation}\label{eq:sigmaL}
\sigma(\bm{n},u) = \sum_{\ell=0}^{+\infty} n_L\,\hat{\sigma}_L(u)\,.
\end{equation}
Notice that the monopole term is given by the angular integral %
\begin{equation}\label{eq:sigma0}
\hat{\sigma}(u) = \int\frac{\dd \Omega}{4\pi}\,\sigma(\bm{n},u) = \frac{4}{c^3 G} \mathcal{F}_\text{GW}\,,
\end{equation}
which agrees with the gravitational-wave energy flux~\eqref{eq:fluxE}.

At this stage it is convenient, following and extending Refs.~\cite{BDLW10b, LBW12}, to perform a gauge transformation in order to 
facilitate the non-linear iteration of the metric and the derivation of the equations of motion and the action. Our choice for the gauge vector is, for the tail part,
\begin{subequations}\label{eq:xi2app}
	\begin{align}
	\xi^{0}_{2}\Big|_\text{tail} &= 4M c^{-4} \log\left(\frac{r}{\lambda}\right) \sum_{\ell=2}^{+\infty} \frac{(-)^\ell}{\ell!}\frac{2\ell+1}{\ell(\ell-1)} \partial_L\!\biggl[\frac{I_L(t-r/c) - I_L(t+r/c)}{r}\biggr]\,, \\
	\xi^{i}_{2}\Big|_\text{tail} &= 4M c^{-5} \log\left(\frac{r}{\lambda}\right) \sum_{\ell=2}^{+\infty} \frac{(-)^\ell}{\ell!}\frac{2\ell-1}{\ell-1} \partial_{L-1}\!\biggl[\frac{I_{iL-1}^{(1)}(t-r/c) - I_{iL-1}^{(1)}(t+r/c)}{r}\biggr]\nonumber\\
	& + 4M c^{-5} \log\left(\frac{r}{\lambda}\right) \sum_{\ell=2}^{+\infty} \frac{(-)^\ell}{\ell!}\frac{2\ell(2\ell+1)}{(\ell+1)(\ell-1)} \varepsilon_{iab}\partial_{aL-1}\!\biggl[\frac{J_{bL-1}(t-r/c) - J_{bL-1}(t+r/c)}{r}\biggr]\,.
	\end{align}
\end{subequations}
For the memory part, we must be careful to define the gauge transformation in such a way that it avoids time anti-derivatives that are incompatible with the conservative dynamics~\cite{LBW12}. Our choice is (with $\hat{\partial}_{iL}\equiv\text{STF}[\partial_{iL}]$)
\begin{subequations}\label{eq:xi2appb}
	\begin{align}
	\xi^{0}_{2}\Big|_\text{mem} &= \frac{1}{4}\log\left(\frac{r}{\lambda}\right)\sum_{\ell=1}^{+\infty} (-)^\ell c^{\ell} \partial_L\!\biggl[\frac{\hat{\sigma}^{(-\ell-2)}_L(t-r/c) - \hat{\sigma}^{(-\ell-2)}_L(t+r/c)}{r}\biggr]\,, \\
	\xi^{i}_{2}\Big|_\text{mem} &= - \frac{1}{4}\log\left(\frac{r}{\lambda}\right)\sum_{\ell=0}^{+\infty} (-)^\ell c^{\ell+1} \hat{\partial}_{iL}\!\biggl[\frac{\hat{\sigma}^{(-\ell-3)}_L(t-r/c) - \hat{\sigma}^{(-\ell-3)}_L(t+r/c)}{r}\biggr]\nonumber\\
	& - \frac{1}{4}\log\left(\frac{r}{\lambda}\right)\sum_{\ell=1}^{+\infty} (-)^\ell c^{\ell-1} \frac{\ell}{\ell+1} \partial_{L-1}\!\biggl[\frac{\hat{\sigma}^{(-\ell-1)}_{iL-1}(t-r/c) - \hat{\sigma}^{(-\ell-1)}_{iL-1}(t+r/c)}{r}\biggr]\,.
	\end{align}
\end{subequations}
After performing this linear gauge transformation the new metric, say ${h'}_2^{\mu\nu} = h_2^{\mu\nu} + \partial^\mu\xi_2^\nu + \partial^\nu\xi_2^\mu - \eta^{\mu\nu}\partial_\rho\xi_2^\rho$, has space components $ij$ that represent a subdominant 0.5PN effect with respect to the components $0i$, themselves 0.5PN smaller than the $00$ component. Finally we find that at the 1PN relative order for mass moments $I_L$ and Newtonian order for current moments $J_L$ the tail part of the metric is 
\begin{subequations}\label{eq:h2primetail}
	\begin{align}
	\bigl({h'}^{00}_{2} + {h'}^{ii}_{2}\bigr)\Big|_\text{tail} &= -16M \log\left(\frac{r}{\lambda}\right)\sum_{\ell=2}^{+\infty} \frac{(-)^\ell a_\ell}{c^{2\ell+6}} \biggl[ \hat{x}_L\,I_L^{(2\ell+2)} + \frac{r^2 \hat{x}_L}{2c^2(2\ell+3)}\,I_L^{(2\ell+4)}\biggr]\,, \\
	{h'}^{0i}_{2}\Big|_\text{tail} &= 8M \log\left(\frac{r}{\lambda}\right)\sum_{\ell=2}^{+\infty} \frac{(-)^\ell}{c^{2\ell+7}} \left[ \frac{a_\ell(2\ell+1)}{(\ell+1)(\ell+2)(2\ell+3)} \,\hat{x}_{iL} \,I_{L}^{(2\ell+3)} - \frac{b_\ell}{2} \,\hat{x}_{aL-1} \varepsilon_{iab}\,J_{bL-1}^{(2\ell+2)}\right]\,,\\
	{h'}^{ij}_{2}\Big|_\text{tail} &= -8M \log\left(\frac{r}{\lambda}\right)\sum_{\ell=2}^{+\infty} \frac{(-)^\ell a_\ell}{c^{2\ell+6}}\biggl[ \frac{2\ell^2 + \ell +1}{(\ell+1)(\ell+2)} \delta_{ij}\, \hat{x}_{L} \,I_{L}^{(2\ell+2)} - \frac{2\ell(2\ell-1)}{(\ell+1)(\ell+2)}\,\hat{x}_{L-1(i} \,I_{j)L-1}^{(2\ell+2)}\biggr]\,,
	\end{align}
\end{subequations}
in which we have conveniently inserted the definitions of the coefficients $a_\ell$ and $b_\ell$ as given by Eq.~\eqref{eq:abell}. The memory part of the metric is drastically simplified in the new gauge as
 \begin{subequations}\label{eq:h2primemem}
 	\begin{align}
 	\bigl({h'}^{00}_{2} + {h'}^{ii}_{2}\bigr)\Big|_\text{mem} &= \frac{1}{c^2}\log\left(\frac{r}{\lambda}\right)\left[ 2 \hat{\sigma} + \frac{r^2}{3c^2} \hat{\sigma}^{(2)}\right]\,, \\
 	{h'}^{0i}_{2}\Big|_\text{mem} &= - \frac{1}{6c^3} \log\left(\frac{r}{\lambda}\right)\left[ x^i \hat{\sigma}^{(1)} - \hat{\sigma}_i\right]\,,\\
 	{h'}^{ij}_{2}\Big|_\text{mem} &= \frac{1}{2c^2} \delta^{ij} \log\left(\frac{r}{\lambda}\right) \hat{\sigma}\,.
 	\end{align}
 \end{subequations}
With the results~\eqref{eq:h2primetail}--\eqref{eq:h2primemem} the components of the ordinary covariant metric $g_{\mu\nu} = \eta_{\mu\nu} + \ell_{\mu\nu}$ are readily computed to quadratic order [we have $\ell_2^{\mu\nu} = -h_2^{\mu\nu} + \frac{1}{2}\eta^{\mu\nu}h_2$ and neglect cubic contributions], and once we have obtained the metric at any point $(\mathbf{x}, t)$ in the near zone, we plug it into the equations of motion of the matter source. In the case of point particles these are simply the geodesic equation which gives the ordinary acceleration $a_1^i=\dd v_1^i/\dd t$ of one of the particle 1 as
\begin{equation}\label{eq:geodeq}
a_1^i = \frac{c^2}{2}\left(1 + \frac{v_1^2}{2c^2}\right)\bigl(\partial_i \ell_{00}\bigr)_1 + c  v_1^j\bigl(\partial_i \ell_{0j}\bigr)_1 + \frac{1}{2} v_1^jv_1^k \bigl(\partial_i \ell_{jk}\bigr)_1 - \frac{\dd}{\dd t}\biggl[\frac{1}{2} v_1^i\bigl(\ell_{00}\bigr)_1 + c \bigl(\ell_{0i}\bigr)_1 + v_1^j\bigl(\ell_{ij}\bigr)_1\biggr]\,.
\end{equation} 
We neglect higher terms in $G$ such as the contribution of the Newtonian potential $U_1 = \frac{G m_2}{r_{12}}$. Similarly, with this approximation the acceleration dependent terms in the right-hand side of~\eqref{eq:geodeq} will be neglected.

In Eq.~\eqref{eq:geodeq} the components of the metric and its gradient are evaluated at the location $y_1^i$ of the particle 1. Thus the logarithms $\ln(r/\lambda)$ in the metric~\eqref{eq:h2primetail}--\eqref{eq:h2primemem} become $\ln(\vert\mathbf{y}_1\vert/\lambda)$ in a general frame, \emph{i.e.} just $\ln(r_{12}/\lambda)$ for binaries in the center-of-mass frame, where we neglect irrelevant constant contributions. Hence the relevant logarithm is $\log(\frac{r_{12}\omega}{c})$ where $\omega$ is the orbital period, or in other words $\log(v/c) = \frac{1}{2} \log\gamma + \text{non-log contributions}$.

To obtain the corresponding action we request that under an arbitrary infinitesimal variation $\delta y_1^i$ of the trajectory of the particle 1 (with the requirement that $\delta y_1^i=0$ when $t=\pm\infty$), holding fixed the trajectories of the other particles, the variation of the action should be 
\begin{equation}\label{eq:deltaS}
\delta S = \int_{-\infty}^{+\infty} \dd t \, m_1 a_1^i \delta y_1^i\,.
\end{equation} 
We insert the metric components~\eqref{eq:h2primetail}--\eqref{eq:h2primemem} into~\eqref{eq:geodeq} and then~\eqref{eq:deltaS}, and perform various manipulations and removal of total time derivatives. In this calculation it is crucial to recognize the 1PN-accurate expression of the mass multipole moments $I_L$ as well as the Newtonian current type ones $J_L$, given by~\cite{BD89, BS89}
\begin{subequations}\label{eq:ILJL}
	\begin{align}
	I_L &= m_1\left[\left(1 + \frac{1}{c^2}\left[\frac{3}{2}v_1^2 - U_1\right]\right) \hat{y}_1^L + \frac{1}{2c^2(2\ell+3)} \frac{\dd^2}{\dd t^2}\left( y_1^2 \hat{y}_1^L\right) - \frac{4(2\ell+1)}{c^2(\ell+1)(2\ell+3)} \frac{\dd}{\dd t}\left( v_1^i \hat{y}_1^{iL}\right)\right] + 1\leftrightarrow 2\,,\\
	J_L &= m_1 \varepsilon_{ab\langle i_\ell} \,\hat{y}_1^{L-1\rangle a} \,v_1^b + 1\leftrightarrow 2\,,
	\end{align} 
\end{subequations}
where we can neglect here the term $U_1$. Finally we obtain for the tail part of the metric, Eq.~\eqref{eq:h2primetail}, after a somewhat involved calculation and still neglecting cubic-order terms $\propto G^3$,
\begin{align}\label{eq:deltaStail}
\delta S\Big|_\text{tail} &= \frac{4 G^2 M}{c^{2\ell+4}}(-)^\ell \int_{-\infty}^{+\infty} \dd t \,\log\left(\frac{r_{12}\omega}{c}\right) \left[ a_\ell \,\delta I_L\,I_L^{(2\ell+2)} + \frac{b_\ell}{c^2} \,\delta J_L\,J_L^{(2\ell+2)}\right]\nonumber\\
&= - \frac{2 G M}{c^{3}} \int_{-\infty}^{+\infty} \dd t \,\log\left(\frac{r_{12}\omega}{c}\right)\delta \mathcal{F}_\text{GW} \,.
\end{align} 
Interestingly, at this stage the total mass $M$ has not been varied yet. The missing piece, implying the variation of the mass, is provided by the memory part of the metric, given by~\eqref{eq:h2primemem}. Indeed we find, using $\delta M = m_1 v_1^i\delta v_1^i + \mathcal{O}(G)$ appropriate to this order, together with the link~\eqref{eq:sigma0} between the angular average $\hat{\sigma}$ and the GW flux (the contribution from $\hat{\sigma}_i$ being higher order), that 
\begin{align}\label{eq:deltaSmem}
\delta S\Big|_\text{mem} &= - \frac{2 G}{c^{3}} \int_{-\infty}^{+\infty} \dd t \,\log\left(\frac{r_{12}\omega}{c}\right) \delta M \,\mathcal{F}_\text{GW}\,,
\end{align} 
so we are finally able to reconstitute the total action in the same form as that for the log-term contributions in Eq.~\eqref{eq:Stail}, see also Eq.~\eqref{eq:Stailcirc}, namely
\begin{align}\label{eq:Sfinal}
S &= -\frac{2 G M}{c^{3}} \int_{-\infty}^{+\infty} \dd t \,\log\left(\frac{r_{12}\omega}{c}\right) \mathcal{F}_\text{GW}\,.
\end{align} 

\section{Derivation of the extra contribution in the non-local energy}
\label{sec:app_nonlocal}
In this Appendix, we present the derivation of the non trivial correction to be added to the conserved energy, in the case of a non-local dynamics. We then show that its time average is directly related to the total energy flux emitted in GWs, as in \eqref{eq:average}. Let us recall that this term satisfies Eq.~\eqref{eq:deltaHdef}
\begin{equation}\label{eq:app_deltaHdef}
\frac{\dd \delta H_\ell}{\dd t} = I_L^{(\ell+1)} \mathcal{I}_L^{(\ell+2)} - I_L^{(\ell+2)} \mathcal{I}_L^{(\ell+1)}\,.
\end{equation}
By writing the functional~\eqref{eq:taildef} as
\begin{equation}\label{eq:tailtau}
\mathcal{I}_L (t) = 
\mathop{\text{Pf}}_{\tau_0} \int_{-\infty}^{+\infty} \frac{\dd \tau}{\vert
  \tau\vert} \,I_{L}(t+  \tau)\,,
\end{equation}
one can perform a formal Taylor expansion of~\eqref{eq:app_deltaHdef} when $\tau \rightarrow 0$. As the kernel $1/\vert\tau\vert$ is even, it only suffices to select even powers of $\tau$ in the expansion. This procedure will naturally lead to divergent integrals (due to the behaviour at $\tau \rightarrow \pm\infty$). To cure those divergences, let us modify the kernel by introducing a regulator $\de^{- \epsilon \vert\tau\vert}$, with an arbitrary parameter $\epsilon > 0$, that will be put to 0 at the end of the computation. Thanks to this regulator, the Hadamard partie finie is no longer needed (the integrals are now convergent at $\tau = 0$). With this modified kernel, and after some integrations by part, it comes as a generalization of Eq.~(3.11) of~\cite{BBBFMb}:
\begin{equation}\label{eq:app_deltaHformal}
\delta H_\ell = 
\sum_{n=1}^{+\infty}\biggl[I_{L}^{(\ell+1)} I_{L}^{(\ell+2n+1)} 
- 2\sum_{s=0}^{n-2}(-)^s I_{L}^{(\ell+s+2)} I_{L}^{(\ell+2n-s)} 
+ (-)^n  \bigl(I_{L}^{(\ell+n+1)}\bigr)^2\biggr]
\!\int_{-\infty}^{+\infty}
\frac{\dd \tau\,e^{- \epsilon \vert\tau\vert}}{\vert\tau\vert}
\frac{\tau^{2n}}{(2n)!}\,.
\end{equation}
To resum the Taylor series, we introduce the Fourier decomposition of the multipoles, following~\cite{ABIQ08tail},
\begin{equation}\label{eq:app_Fourier}
I_{L}(t) =
\sum_{p=-\infty}^{+\infty}\sum_{m=-\ell}^{+\ell}\,\mathop{{\tilde{I}}}_{(p,m)}{}_{\!\!L}\,\de^{\di \,(p+m\,\text{k})\,\mathcal{M}} \,,
\quad\text{with}\quad 
\mathop{{\tilde{I}}}_{(p,m)}{}_{\!\!L} =
\int_0^{2\pi}\frac{\dd\mathcal{M}}{2\pi}\,I_{L}\,\de^{-\di\, (p+m\,\text{k})\,\mathcal{M}}\,,
\end{equation}
where $\mathcal{M}=\omega(t-t_0)$ is the mean anomaly of the binary motion, with $\omega=2\pi/P$ being the orbital frequency (or mean motion) corresponding to the orbital period $P$, and $t_0$ an instant of reference. The index $p$ corresponds to the usual orbital motion, and the ``magnetic-type'' index $m$, to the relativistic precession (k being defined by the precession of the periastron \emph{per} period, $\Delta\phi=2\pi\text{k}$). The discrete Fourier coefficients ${}_{(p,m)}{\tilde{I}}_{L}$ naturally satisfy ${}_{(-p,-m)}{\tilde{I}}_{L} = {}_{(p,m)}{\tilde{I}^\star}_{L}$, with the star denoting the complex conjugation. In the following, we will denote $\tilde{p} = p + m \,\text{k}$ and thus ${}_{\tilde{p}}{\tilde{I}}_{L}={}_{(p,m)}{\tilde{I}}_{L}$. Separating the constant (DC) part $\tilde{p}+\tilde{q}=0$ from the oscillating (AC) one, and resumming with respect to $n$, the equation~\eqref{eq:app_deltaHformal} reads
\begin{align}\label{eq:app_deltaHFourier}
\delta H_\ell = & -
\sum_{\tilde{p}}\left(\tilde{p}\omega\right)^{2\ell+2}\,\vert\mathop{{\tilde{I}}}_{\tilde{p}}{}_{\!\!L}\vert^2 
\int_{-\infty}^{+\infty}\frac{\dd \tau\,\de^{-\epsilon\vert\tau\vert}}{\vert\tau\vert} \tilde{p}\omega\tau\,\sin\left(\tilde{p}\omega\tau\right)\nonumber\\
& - \frac{(-)^\ell}{2}\sum_{\tilde{p}+\tilde{q}\neq 0}\left(\tilde{p}\tilde{q}\omega^2\right)^{\ell+1}
\mathop{{\tilde{I}}}_{\tilde{p}}{}_{\!\!L}\mathop{{\tilde{I}}}_{\tilde{q}}{}_{\!\!L}\,\frac{\tilde{p}-\tilde{q}}{\tilde{p}+\tilde{q}}\,\de^{\di \left(\tilde{p}+\tilde{q}\right)\mathcal{M}}
\int_{-\infty}^{+\infty}\frac{\dd \tau\,\de^{-\epsilon\vert\tau\vert}}{\vert\tau\vert} \bigl[\cos\left(\tilde{p}\omega\tau\right)-\cos\left(\tilde{q}\omega\tau\right)\bigr]\,.
\end{align}
Integrating over $\tau$ and setting the regulator $\epsilon$ to zero gives the final formula for this contribution, namely
\begin{equation}\label{eq:app_deltaHfinal}
\delta H_\ell = -2\omega^{2\ell+2}
\biggl[\sum_{\tilde{p}}\,\vert\mathop{{\tilde{I}}}_{\tilde{p}}{}_{\!\!L}\vert^2 \tilde{p}^{2\ell+2}
- \frac{(-)^\ell}{2} \sum_{\tilde{p}+\tilde{q}\neq 0}\,\mathop{{\tilde{I}}}_{\tilde{p}}{}_{\!\!L}\mathop{{\tilde{I}}}_{\tilde{q}}{}_{\!\!L}
  \,\frac{\tilde{p}^{\ell+1}\tilde{q}^{\ell+1}(\tilde{p}-\tilde{q})}{\tilde{p}+\tilde{q}}  \ln\left|\frac{\tilde{p}}{\tilde{q}}\right|\,\de^{\di(\tilde{p}+\tilde{q})\mathcal{M}}\biggr]\,,
\end{equation}
where we recall that $\tilde{p} = p+m\,\text{k}$, with $m\in [-\ell,\ell]$ and k is the relativistic precession. 

Let us define the total contribution in the conserved energy that is due to those extra terms as
\begin{equation}\label{eq:app_deltaHtot}
\delta H = \sum_{\ell=2}^\infty \frac{G^2 M \,a_\ell}{c^{2\ell+4}}\,\delta H_\ell + \text{(current moments contribution)}\,.
\end{equation}
Averaging over one period, only the first term in~\eqref{eq:app_deltaHfinal} survives, and yields
\begin{equation}\label{eq:app_average}
\langle\delta H\rangle = - \frac{2G M}{c^3}\,\sum_{\ell=2}^{+\infty} \frac{G}{c^{2\ell+1}} \left[ a_\ell \,I_L^{(\ell+1)}\,I_L^{(\ell+1)} + \frac{b_\ell}{c^2} J_L^{(\ell+1)}\,J_L^{(\ell+1)} \right]\,.
\end{equation}
This confirms the previously stated relation~\eqref{eq:average}.

\end{document}